\documentclass[fleqn,usenatbib]{mnras}

\usepackage{newtxtext,newtxmath}

\usepackage[T1]{fontenc}

\DeclareRobustCommand{\VAN}[3]{#2}
\let\VANthebibliography\thebibliography
\def\thebibliography{\DeclareRobustCommand{\VAN}[3]{##3}\VANthebibliography}

\usepackage{graphicx}	
\usepackage{amsmath}	
\usepackage{amssymb}	

\title[Striped Blazar Jet]{Radiation Signatures From Striped Blazar Jet}

\author[H. Zhang \& D. Giannios]{
Haocheng Zhang$^{1}$\thanks{E-mail: astrophyszhc@hotmail.com},
Dimitrios Giannios$^{1}$
\\
$^{1}$ Department of Physics, Purdue University, West Lafayette, IN 47907, USA
}

\date{Accepted XXX. Received YYY; in original form ZZZ}

\pubyear{2020}

\begin{document}
\label{firstpage}
\pagerange{\pageref{firstpage}--\pageref{lastpage}}
\maketitle

\begin{abstract}
Relativistic jets from supermassive black holes are among the most powerful and luminous astrophysical systems in Universe. We propose that the open magnetic field lines through the black hole, which drive a strongly magnetized jet, may have their polarity reversing over time scales related to the growth of the magneto-rotational dynamo in the disc, resulting in dissipative structures in the jet characterized by reversing toroidal field polarities, referred to as ``stripes''. Magnetic reconnection between the stripes dissipates the magnetic energy and powers jet acceleration. The striped jet model can explain the jet acceleration, large-scale jet emission, and blazar emission signatures consistently in a unified physical picture. Specifically, we find that the jet accelerates to the bulk Lorentz factor $\Gamma \gtrsim 10$ within one parsec distance from the central engine. The acceleration slows down but continues at larger distances, with intrinsic acceleration rate $\dot{\Gamma}/\Gamma$ between $0.0005~\rm{yr^{-1}}$ and $0.005~\rm{yr^{-1}}$ at tens of parsecs, which is in very good agreement with recent radio observations. Magnetic reconnection continuously accelerates nonthermal particles over large distances from the central engine, resulting in the core-shift effect and overall flat-to-inverted synchrotron spectrum. The large-scale spectral luminosity peak $\nu_{\rm peak}$ is anti-proportional to the location of the peak of the dissipation, which is set by the minimal stripe width $l_{\rm min}$. The blazar zone is approximately at the same location. At this distance, the jet is moderately magnetized, with the comoving magnetic field strength and dissipation power consistent with typical leptonic blazar model parameters.
\end{abstract}




\section{Introduction}
\label{sec:intro}

Relativistic jets from active galactic nuclei (AGNs) are among the most powerful astrophysical systems in Universe. They are believed to originate from strongly accreting supermassive black holes, with initially very high magnetic energy at the launching site. These highly collimated plasma jets can extend to kilo-parsec or even up to Mega-parsec from their central engine \citep{Leahy84}. During the jet propagation, the jet releases enormous amount of radiation energy in the form of nonthermal-dominated emission. In particular, blazars, AGN jets that point very close to our line of sight, exhibit highly variable emission across the entire electromagnetic spectrum, up to TeV $\gamma$-rays \citep{Aharonian2007,Ackermann2016}. This indicates very efficient particle acceleration in a very localized region, often referred to as the ``blazar zone''. Blazars have long been considered as the potential origin of extragalactic cosmic rays and neutrinos \citep{Mannheim1993,Muecke2003,Ojha2019}. The recent blazar flare coincident with a very high energy neutrino event provides hints on the hadronic processes in blazars \citep{IceCube2018}. However, the blazar jet involves very complicated physical processes over many orders of physical scales \citep[see e.g.,][]{Rani2019,Boettcher2019}. Therefore, it is very important to have a unified physical picture that can explain the observed jet kinematics, large-scale emission, and the blazar zone radiation signatures consistently.


Jet kinematics and spatially resolved radiation signatures can diagnose the overall jet physical conditions and energy evolution. Very long baseline interferometry (VLBI) techniques have been widely used to explore the AGN jet structures with very high spatial resolution. Radio monitoring programs such as MOJAVE and BU-BLAZAR projects have revealed rich spatially resolved features close to the central engine in many blazars \citep{Marscher2008,Jorstad2017,Lister2018}, such as bright compact radio cores near the jet base, as well as standing and moving radio knots along the jet propagation direction. The moving radio knots can appear to travel faster than the speed of light, known as the apparent superluminal motion, showing that the jet is moving at relativistic speeds \citep{Rees1966}. The overall radio spectrum appears flat or even slightly inverted, and the radio core appears more compact in higher-frequency bands (often called the core-shift effect). \citet{Blandford1979} have successfully interpreted these radio phenomena as synchrotron emission and self absorption along the jet propagation from a conical jet. This model relies on unspecified mechanisms of continuous jet energy dissipation to accelerate nonthermal particles across a very broad range of length scales along the jet propagation. But the physical origin of the continuous jet energy dissipation is so far not well understood.

The bright and strongly variable emission from the blazar zone also poses a major theoretical challenge to efficiently dissipate jet energy and accelerate nonthermal particles in very localized regions. Blazars are the most numerous extragalactic $\gamma$-ray sources observed by {\it Fermi} \citep{Abdollahi2020}, due to the efficient particle acceleration in the blazar zone and relativistic beaming effects. All wavelengths show considerable variability in flux, and many observations have shown flaring time scales between days and weeks \citep{Abdo2010,Sandrinelli2014,Rajput2020}. The causality relation then requires the size of the blazar zone to be of sub-parsec size. In addition to the flux variability, recent optical polarization observations have witnessed simultaneous polarization variability, including large angle swings, with multi-wavelength blazar flares \citep{Marscher2008,Chandra2015,Blinov2018}. These observations imply active magnetic field evolution along with particle acceleration \citep{Larionov2013,Zhang2015}. The location of the blazar zone is often believed to be at sub-parsec to parsec scales from the central engine, but its physical conditions and how it efficiently dissipates so much jet energy to accelerate nonthermal particles have been long-standing questions to both observations and theories.

An interesting recent discovery is that the jet is still accelerating at tens of parsecs from the central engine. Recent VLBI monitoring programs have been able to study the apparent acceleration of large numbers of individual jet features across many jets \citep{Jorstad2017,Lister2018}. For instance, MOJAVE project has revealed significant parallel acceleration of jet features at projected distances of parsec scales in many jets \citep{Homan2015}, indicating jet bulk acceleration at tens of parsecs from the central engine. These observations indicate that the jet magnetic energy can continue to be converted to bulk kinetic energy beyond the blazar zone.


This paper aims to address the above issues theoretically under the striped jet model. This model envisions that the magneto-rotational instabilities (MRI) that drive the black hole accretion processes generate loops of magnetic field of varying polarity \citep{Balbus1991}. Due to the black hole rotation, these magnetic loops are forced to open up rapidly. The rotating black hole threaded by open magnetic field lines then drives a highly magnetized jet \citep{Uzdensky2005,Barkov2011,Parfrey2015}. The varying magnetic field polarity through the black hole results in a jet characterized by a varying toroidal field polarity as followed along its propagation axis. The narrow segments of uniform polarity of magnetic field are, hereafter, referred to  as "stripes". The jet acceleration and energy dissipation during its propagation are then driven by magnetic reconnection between stripes of opposite polarity. The dynamics of such a striped jet have been studied by \citet{Drenkhahn2002a} and \citet{Drenkhahn2002b} in the context of GRB jets, assuming a single unique stripe width throughout the jet. Recently, \citet{Giannios2019} have considered a generalized striped jet model, where the stripe widths follow a broad distribution determined by the temporal power spectrum of MRI-driven dynamo at different scales in the accretion disc. Under this model, \citet{Giannios2019} have shown that the jet bulk acceleration and energy dissipation cover a broad range of distances from the jet central engine, depending on the stripe distribution profile. The model makes specific predictions for the energy dissipation and bulk acceleration profile of the jet and, therefore, allows for the development of a unified physical picture to explain the blazar jet dynamics and radiation signatures.


Here we study the blazar jet kinematics, large-scale radio signatures, and blazar zone radiation features following the generalized striped jet model by \citet{Giannios2019}. We explore the dependence of observable signatures on the striped jet model parameters, and whether we can reproduce the typical observable signatures with reasonable physical parameters. We investigate the jet bulk acceleration, especially at tens of parsecs, and compare with MOJAVE observations. We also evaluate the synchrotron emission and self absorption signatures along the jet propagation direction following \citet{Blandford1979}, and explore core-shift effects and overall radio spectra. We try to locate the blazar zone in the striped jet and figure out its physical conditions. Then we study if these physical parameters can reproduce the typical blazar spectra and variability patterns. Most importantly, we show how the striped jet model mutually links the above three independent aspects of observable signatures, i.e., the large-scale jet acceleration, radio emission, as well as the properties and location of the blazar zone, and connects them with the physical processes at the central engine. The paper is organized as follows. Section \ref{sec:model} reviews the key assumptions and results of the striped jet model based on \citet{Giannios2019}; Section \ref{sec:kinematics} studies the jet acceleration and dissipation during its propagation; based on our radiation model, Section \ref{sec:radiojet} studies the core-shift effects and radio spectra of the striped jet; Section \ref{sec:blazarzone} explores the physical conditions and radiation signatures of the blazar zone. We summarize and discuss our results in Section \ref{sec:discussion}.

\section{Striped Jet Model}
\label{sec:model}

In this section, we review the key assumptions and results of the striped jet model based on \citet{Giannios2019}. As we can see below, the striped jet dynamics and evolution mostly depend on the stripe distribution profile. This profile is characterized by two key parameters, namely, the minimal stripe width $l_{\rm min}$ and the power spectral index $a$. Both parameters are directly related to the black hole accretion and jet launching processes.

It is believed that the black hole accretion processes are driven by the MRI in the accretion disc \citep{Balbus1991}. In this picture, the MRI-driven dynamo generates magnetic field loops of varying polarity. The accreting gas advects the inner footpoints of the magnetic loops into the black hole while the outer footpoints still thread the disc. The resulting differential rotation forces the magnetic field loops to open up rapidly \citep{Uzdensky2005,Parfrey2015,Barkov2011}. A rotating black hole threaded by these open magnetic field lines then drives a strongly magnetized jet \citep{Blandford1977}. The polarity of the MRI-driven magnetic field loops can vary over time-scales comparable to the orbital time in the inner parts of the disc. Therefore, the polarity of the open magnetic field lines through the black hole can reverse over a range of time-scales related to the characteristic MRI growth time in the disc, which in turn scales with the Keplerian period. The resulting jet is then characterized by varying toroidal field polarity along its propagation axis, which we refer to these structures as stripes.

Recent magnetohydrodynamic (MHD) simulations of accretion-disc turbulence driven by the MRI show that MRI dynamo in stratified discs produces quasi-periodic reversals of the magnetic flux that is emerging vertically out from the disc into the wind and jet \citep{Davis2010,Oneill2011,Simon2012}. The characteristic period for these reversals is about ten Keplerian orbital periods $T_{\rm orb}=(R/R_{\rm g})^{3/2}R_{\rm g}/c$ at the distance of interest $R$, where $R_{\rm g}=GM/c^2$ is the gravitational radius. A significant amount of the jet energy comes from the inner part of the accretion disc, with a radius of a few to tens of gravitational radii. Therefore, the minimal stripe width $l_{\rm min}\sim 10 cT_{\rm orb}$ can be estimated to be hundreds to thousands of $R_{\rm g}$, depending on how fast the black hole is rotating.\footnote{The black hole spin sets the location of the inner edge of the disc and, therefore, the shortest Keplerian period timescale of the disc.} In a more realistic picture, MRI turbulence is active over a broad range of radii in the disc as suggested by recent shearing box and GRMHD simulations, thus introducing magnetic field reversals over a broad range of scales \citep{Ball2018,Kadowaki2018,Kadowaki2019}. However, for relatively small scale turbulence, the magnetic reversals can quickly dissipate via magnetic reconnection near the base of the jet or inside the accretion disk. Therefore, we expect that their contribution to the global jet dynamics and propagation should be small. To quantitatively study the contribution of small scale magnetic field reversals in the jet, one can examine the ratio of turbulent magnetic field component over the coherent magnetic field reversals numerically via GRMHD of the jet launching processes, which is beyond the scope of this paper. For simplicity, here we only consider the coherent magnetic field reversals and take a truncated power-law distribution of the stripes from $l_{min}$ as suggested by \citet{Giannios2019},
\begin{equation}
\mathcal{P}(l)\equiv\frac{dP}{dl}=\frac{1}{(a-1)l_{\rm min}}
\left(\frac{l}{l_{\rm min}}\right)^{-a}~~,~~l\ge l_{\rm min}~~.
\end{equation}
In this way, the stripe distribution profile at the jet launching site is characterized by two parameters, $l_{\rm min}$ and $a$, both of which are related to the black hole accretion processes. As discussed above, $l_{\rm min}\sim 10^2-10^3 R_{\rm g}$ may be a reasonable estimate. The $a$ parameter is, theoretically, poorly constrained. In general, $a>1$ appears to be a natural assumption, since most of the gravitational energy is released at the inner disc, implying that the energetically dominant scale of the stripes is $l_{\rm min}$. If the stripe distribution follows that of gravitational energy release, then $a=5/3$ \citep{Giannios2019}.

It is the evolution of the stripes during the jet propagation that governs the jet acceleration and magnetic energy dissipation. Initially, close to the jet base, the expansion time of the jet is much shorter than the time it would take for any stripes to dissipate they energy via magnetic reconnection. As a result, the jet is initially not very dissipative. As the jet propagates outwards, it can expand sideways depending on the external pressure profile. For simplicity, we assume that the jet is conical after its initial collimation and acceleration near the jet base, with a fixed small opening angle $\theta_{\rm j}$ of a few degrees. The embedded stripes thus stretch sideways, making them prone to tearing instabilities. At the same time, the jet takes longer to expand leaving plenty of time for efficient magnetic reconnection between stripes of small length scales. At larger distances, longer stripes also have time to dissipate. 

Magnetic dissipation drives the jet bulk acceleration but also injection of nonhtermal particles in the jet. To make the problem analytically tractable,  
 \citet{Giannios2019} have postulated that the jet remains cold throughout the jet propagation and ignored radiative losses. \citet{Drenkhahn2002b} have shown that when the radiative losses are small, the thermal energy of the plasma in the jet always remains a modest fraction of the bulk kinetic energy. Given that the blazar radiation efficiency is typically about ten percent of the jet power, radiative losses do not significantly affect the dissipation and acceleration of the jet.

Under the above assumptions, the total jet power $L_{\rm j}$ is conserved during the jet propagation. At the jet base, it is mostly in the form of Poynting power $L_{\rm p}$. During the jet propagation, the jet magnetic energy continuously converts into the bulk kinetic energy $L_{\rm k}$. The jet reaches the terminal Lorentz factor $\Gamma_{\rm \infty}=L_{\rm j}/(\dot{M}c^2)$ very far from the central engine, where $\dot{M}$ is the mass injection rate at the jet base. Thus we have
\begin{equation}
L_{\rm j}\equiv L_{\rm k}+L_{\rm p}=L_{\rm k}(1+\sigma)=\Gamma \dot{M}c^2(1+\sigma)~~,
\end{equation}
where $\sigma=L_{\rm p}/L_{\rm k}$ is the magnetization factor. For a relativistic jet, \citet{Giannios2019} have worked out the differential equation that describes the jet acceleration during its propagation,
\begin{equation}
\frac{d\chi}{d\zeta}=\frac{(1-\chi)^k}{\chi^2}~~,
\label{eq:acceleration}
\end{equation}
where $\chi=\Gamma/\Gamma_{\rm \infty}$, $\zeta=(2\epsilon z)/(l_{\rm min}\Gamma^2_{\rm \infty})$, $z$ is the distance from the central engine, and $k=(3a-1)/(2a-2)$. Here $\epsilon$ is the reconnection rate measure as a fraction of the Alfv\'en velocity. Particle-in-cell (PIC) simulations have found that in the relativistic magnetic reconnection, $\epsilon\sim 0.1$ generally holds for a magnetized ($\sigma \gtrsim 1$) plasma environment \citep{Liu2017}. The above equation can be solved for an implicit expression of $\chi(\zeta)$,
\begin{equation}
\frac{(1-\chi)^{3-k}}{k-3}-\frac{2(1-\chi)^{2-k}}{k-2}+\frac{(1-\chi)^{1-k}}{k-1}=\zeta+C ~~,
\label{eq:sj_eqn}
\end{equation}
if $k\ne 1,\,2,\,3$, where $C$ is a constant. If $k=2$ or $k=3$ (for $a>1$, $k$ cannot reach 1), the corresponding term is replaced by a logarithmic term. Under the assumption that the thermal energy and radiative losses are neglected, \citet{Giannios2019} have shown that the jet magnetic energy dissipation at a specific distance from the central engine can be considered as the change of the bulk kinetic energy there. 
In this way, the jet dissipation power $P_{\rm diss}$ is given by
\begin{equation}
P_{\rm diss}=z\frac{dL_{\rm k}}{dz}=L_{\rm j}\frac{(1-\chi)^k}{\chi^2}\zeta~~.
\label{eq:dissipation}
\end{equation}
\citet{Giannios2019} have shown that the jet acceleration and dissipation are continuous over several orders of distances from the central engine. Additionally, the dissipation peaks at approximately
\begin{equation}
z_{\rm peak}=\frac{l_{\rm min}\Gamma^2_{\rm \infty}}{6\epsilon}~~.
\end{equation}
The peak distance corresponds to the location where the smallest stripes $l_{\rm min}$ mostly dissipate.

We summarize the key assumptions and results from \citet{Giannios2019}:
\begin{enumerate}
\item The jet is conical.
\item At the jet launching site, the jet magnetic field is predominantly toroidal, in the form of stripes of reversing polarity.
\item The stripes are separated by current sheets where their magnetic energy is dissipated via magnetic reconnection.
\item The jet remains cold throughout its propagation.
\item The radiative losses are small.
\item Under the above assumptions, the jet acceleration and dissipation during its propagation are described by Equations \ref{eq:acceleration} and \ref{eq:dissipation}, respectively. Both change slowly over several orders of magnitude in distance from the central engine.
\end{enumerate}

\section{Jet Kinematics and Dissipation}
\label{sec:kinematics}

Under the striped jet model, the jet acceleration is continuous over a broad range of distances from the central engine. The acceleration is relatively fast near the base of the jet, and the bulk Lorentz factor reaches to $\gtrsim 10$ within $\sim 1~\rm{pc}$. When the distance $z_{\rm peak}$ is reached, the jet acceleration slows down, but continues to ten to a hundred parsecs from the central engine, with the intrinsic acceleration rate $\dot{\Gamma}/\Gamma$ between $0.0005~\rm{yr^{-1}}$ to $0.005~\rm{yr^{-1}}$. These values are in good agreement with the MOJAVE observations \citep{Homan2015}. The acceleration is driven by conversion of the jet Poynting flux into bulk kinetic flux via magnetic reconnection. This is because the jet is still moderately magnetized beyond the blazar zone, and only depletes most of its magnetic energy at $\gtrsim 100~\rm{pc}$.

\subsection{Overall Jet Acceleration and Dissipation}
\label{sec:acceleration}

\begin{figure}
\centering
\includegraphics[width=\linewidth]{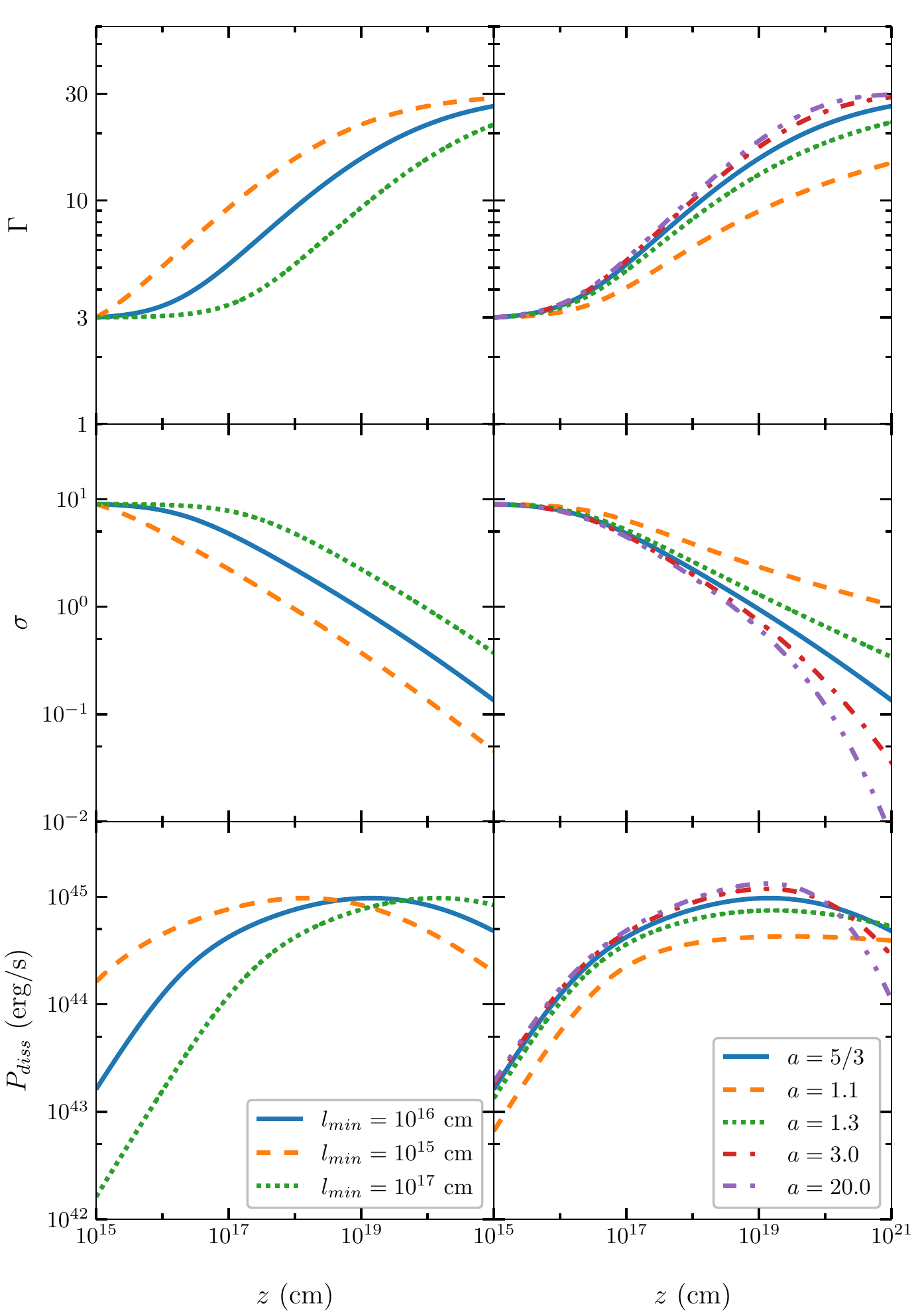}
\caption{The jet bulk Lorentz factor $\Gamma$ (top panels), magnetization factor $\sigma$ (middle panels), and dissipation power $P_{\rm diss}$ (bottom panel) along the jet propagation direction. The left panels show the dependence on the minimal stripe width $l_{\rm min}$, while the right panels are for different stripe profile index $a$.}
\label{fig:evolution}
\end{figure}

The striped jet model predicts jet bulk acceleration over a broad range of distances from the central engine. Equation \ref{eq:acceleration} suggests that the jet acceleration is fast near the base of the jet, where the jet is relatively slow. The acceleration quickly drops when the jet bulk Lorentz factor approaches the terminal Lorentz factor $\Gamma_{\rm \infty}$. As shown in \citet{Giannios2019}, we can find the asymptotic expressions for $\chi \ll 1$ and $\chi \rightarrow 1$, which are
\begin{equation}
\begin{aligned}
& \chi^3 =3\zeta ~~, &\chi \ll 1~~, \\
& (1-\chi)^{1-k} =(k-1)\zeta ~~, &\chi \rightarrow 1~~.
\end{aligned}
\label{eq:asymptotic}
\end{equation}
The jet bulk Lorentz factor at small distances from the central engine thus can be expressed as
\begin{equation}
\Gamma=(6\epsilon \Gamma_{\rm \infty})^{\frac{1}{3}}\Big(\frac{z}{l_{\rm min}}\Big)^{\frac{1}{3}} \sim 15 \Big(\frac{\Gamma_{\rm \infty}}{30}\Big)^{\frac{1}{3}}\Big(\frac{l_{\rm min}}{10^{16}~\rm{cm}}\Big)^{-\frac{1}{3}}\Big(\frac{z}{1~\rm{pc}}\Big)^{\frac{1}{3}} ~~,
\label{eq:smalllorentz}
\end{equation}
where we consider a jet with terminal Lorentz factor of $\Gamma_{\rm \infty}=30$ and the minimal striped width $l_{\rm min}=10^3R_{\rm g}\sim 10^{16}~\rm{cm}$ for a supermassive black hole of $10^8$ solar masses.\footnote{Hereafter, when we convert distance from $R_{\rm g}$ units to cm, we assume a black hole of $10^8$ solar masses, unless stated otherwise.} One can see that the bulk jet Lorentz factor reaches $\gtrsim 10$ within $\sim 1~\rm{pc}$, which is consistent with observations. The acceleration is gradual, as shown in the factor $(z/l_{\rm min})^{1/3}$. Apparently, the acceleration at small distances from the central engine does not depend on the stripe spectral index $a$, but depends on the minimal stripe width $l_{\rm min}$. The physical reason is that after the initial jet acceleration and collimation near the central engine, the jet acceleration is dominated by the dissipation of the narrowest stripes. Aforementioned, the jet radial expansion stretches the stripes sideways, making them unstable to tearing instabilities. This happens at smaller distance $z$ from the central engine for narrower stripes. Therefore, the initial jet dissipation is dominated by magnetic reconnection in stripes of width $\sim l_{\rm min}$, regardless of the stripe index $a$. If the central black hole rotation is fast, which results in smaller $l_{\rm min}$, the jet bulk acceleration and dissipation can be much faster.

Figure \ref{fig:evolution} (upper row) shows the jet bulk Lorentz factor profile solved numerically based on Equation \ref{eq:sj_eqn}. We can see that the profile follows well with the above analysis. For different $l_{\rm min}$, the Lorentz factor profile shifts horizontally in $z$, which is expected from the $(z/l_{\rm min})^{1/3}$ factor in Equation \ref{eq:smalllorentz}. And the acceleration near the jet base is similar for different $a$. Generally speaking, for larger $a$, there are more stripes of width $\sim l_{\rm min}$, thus we see slightly faster acceleration. The jet magnetization (Figure \ref{fig:evolution} middle row) follows the opposite trend to the bulk Lorentz factor. This is expected, as the jet bulk kinetic energy comes from the conversion of the Poynting flux via reconnection in stripes.

Interestingly, we see that both the Lorentz factor and magnetization factor show a broad linear profile in the logarithmic space, covering several orders of magnitude in distance from the central engine. Figure \ref{fig:evolution} (lower row) shows that the dissipation power exhibits a broad peak there. \citet{Giannios2019} have shown that this broad peak around $z_{\rm peak}$ marks the location where the narrowest stripes ($l_{\rm min}$) strongly dissipate. As a result, the jet magnetization quickly drops from nearly ten to one. Since the dissipation before $z_{\rm peak}$ is dominated by the dissipation of narrowest stripes, the dissipation profile appears similar for different $a$ before $z_{\rm peak}$. After $z_{\rm peak}$, the narrowest stripes have been dissipated. Thus for smaller $a$, which means more stripes of larger width, the dissipation  continues to longer distances. Otherwise the dissipation power quickly drops beyond $z_{\rm peak}$. Since $z_{\rm peak}$ is proportional to $l_{\rm min}$, we also see that the profile shifts horizontally in $z$ with different $l_{\rm min}$. This peak dissipation region is likely the blazar zone under the striped jet model. We can see that the jet still has a considerable amount of the magnetic energy ($\sigma >0.1$) even at $\sim 10~\rm{pc}$ from the central engine (see more discussions in Section \ref{sec:location}). This drives further acceleration beyond the blazar zone.

\subsection{Jet Acceleration Beyond Blazar Zone}
\label{sec:accelerationlong}

\begin{figure}
\centering
\includegraphics[width=\linewidth]{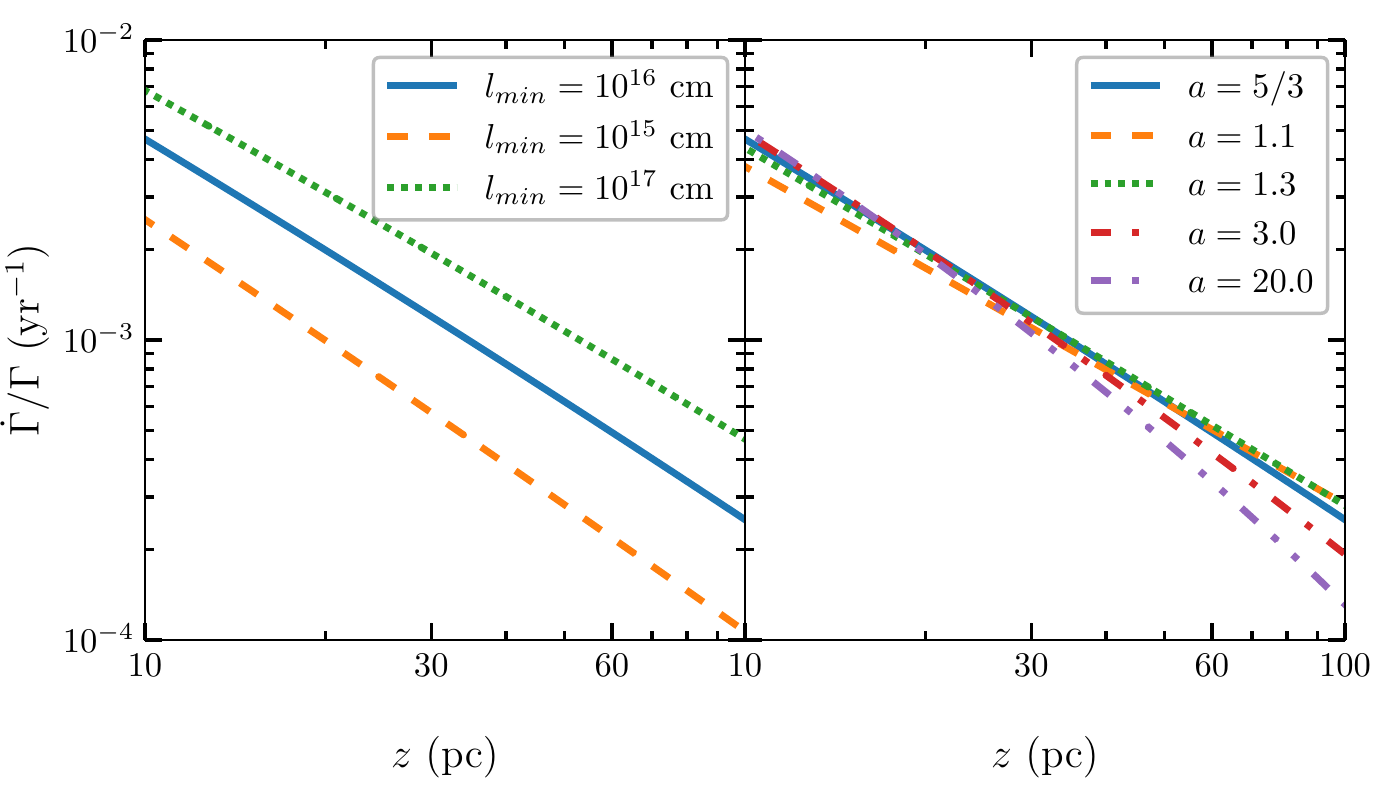}
\caption{The jet intrinsic acceleration rate $\dot{\Gamma}/\Gamma$ at ten to one hundred parsecs from the central engine. The left panels show the dependence on the minimal stripe width $l_{\rm min}$, while the right panels are for different stripe profile index $a$.}
\label{fig:acceleration}
\end{figure}

Very interestingly, the jet bulk acceleration continues beyond the peak dissipation (blazar zone). We can see in Figure \ref{fig:evolution} (upper row) that the bulk Lorentz factor increases from $\sim 15$ to nearly 30 at tens of parsecs. The acceleration rate can be expressed as
\begin{equation}
\frac{\dot{\Gamma}}{\Gamma}=\frac{c}{\Gamma}\frac{d\Gamma}{dz} ~~.
\end{equation}
At tens of parsecs, the jet has moved out the linear acceleration phase in logarithmic space, making it hard to estimate the acceleration rate analytically. Figure \ref{fig:acceleration} solves it numerically from 10 to 100 parsecs. We can see that the acceleration rate gradually drops with longer distances. The acceleration rate has a stronger dependence on $l_{\rm min}$ than $a$. This is due to the cutoff of the dissipation power beyond $z_{\rm peak}$ as shown in Figure \ref{fig:evolution}. We can see that the magnetization drops below 1 after $z_{\rm peak}$, so that the dissipation via magnetic reconnection becomes inefficient. As a result, the acceleration drops after $z_{\rm peak}$. Nevertheless, the acceleration at tens of parsecs is still considerable given the large spatial scale. Since $z_{\rm peak}$ is proportional to $l_{\rm min}$, the acceleration rate strongly depends on $l_{\rm min}$.

The acceleration rate under the striped jet model is consistent with recent VLBI observations. \citet{Homan2015} have found that a large number of radio jets show significant parallel acceleration. Assuming a constant Doppler factor of 10, they have estimated that the intrinsic acceleration rate is between $10^{-3}~\rm{yr^{-1}}$ to $10^{-2}~\rm{yr^{-1}}$ at tens of parsecs. The striped jet model predicts slightly larger bulk Lorentz factor, at $\sim 15$ at tens of parsecs. Thus the observational results yield $5\times 10^{-4}~\rm{yr^{-1}}$ to $5\times 10^{-3}~\rm{yr^{-1}}$. Based on the above constraints, we find that for $l_{\rm min}\gtrsim 10^{16}~\rm{cm}$ and $1.1\lesssim a \lesssim 3$, the striped jet model predicts similar parallel acceleration as in \citet{Homan2015}.

\section{Large-Scale Jet Radiation}
\label{sec:radiojet}

The striped jet model naturally predicts a flat dissipation profile of energy that extends for several orders of magnitude in distance. This property of the dissipation profile explains the core-shift effect and flat-to-inverted radio spectrum as seen in observations. The core-shift distance $z_{\rm cs}(\nu)$ is anti-proportional to the observational frequency $\nu$ for typical striped jet model parameters. Very interestingly, the peak of the radio spectrum $\nu_{\rm peak}$ is anti-proportional to the peak of the dissipation profile $z_{\rm peak}$. Since $z_{\rm peak}$ is proportional to $l_{\rm min}$, which characterizes the creation of stripes via the MRI, the radio spectral peak can be an observable diagnostic of the black hole accretion. In the following, we first describe our radiation model, then evaluate the core-shift relation and large-scale synchrotron spectra.

\subsection{Radiation Model}
\label{sec:radiation}

The striped jet dissipates the jet magnetic energy via magnetic reconnection between stripes of opposite polarity. Numerous PIC simulations of relativistic magnetic reconnection have shown that the reconnection can accelerate particles into a power-law distribution \citep{Sironi2014,Guo2014,Guo2016,Petropoulou2016}. Therefore, we inject a straight power-law distribution of nonthermal particles along the jet propagation direction, whose energy is a fraction of the local dissipation power. The large-scale jet emission is dominated by nonthermal electron synchrotron emission. Given the magnetic field and nonthermal particle distributions based on the striped jet model, we can evaluate the synchrotron emission and synchrotron self absorption from blazars. In the following, all primed quantities are in the comoving frame of the local jet segment. 

Following \citet{Boettcher2012}, the synchrotron emissivity is given by
\begin{equation}
j'_{\rm \nu'}=\frac{1}{4\pi}\int_{ 1}^{\infty}d\gamma'~n'(\gamma')P'_{\rm \nu'}(\gamma') ~~,
\end{equation}
where $n'(\gamma')$ is the nonthermal electron distribution, and $P'_{\rm \nu'}(\gamma')$ is the specific synchrotron power at frequency $\nu'$ by an electron of Lorentz factor $\gamma'$. The synchrotron self absorption coefficient is given by
\begin{equation}
\alpha'_{\rm \nu'}=-\frac{1}{8\pi m_{\rm e}\nu^{\prime\,2}}\int_{ 1}^{\infty}d\gamma'~P'_{\rm \nu'}(\gamma')\gamma^{\prime\,2}\frac{\partial}{\partial \gamma'}\Big(\frac{n'(\gamma')}{\gamma^{\prime\,2}}\Big) ~~.
\label{eq:ssa}
\end{equation}
For isotropically distributed nonthermal particles, the specific synchrotron power is
\begin{equation}
P'_{\rm \nu'}(\gamma')=\frac{\sqrt{3}\pi e^3B'}{2m_{\rm e}c^2}xCS(x)~~,
\label{eq:synpower}
\end{equation}
where $x=\nu'/\nu'_{\rm c}$, $\nu'_{\rm c}=(3eB'\gamma^{\prime\,2})/(4\pi m_{\rm e}c)$ is the synchrotron critical frequency, and the function $CS(x)$ is given in terms of Whittaker's functions $CS(x)=W_{\rm 0,\frac{4}{3}}(x)W_{\rm 0,\frac{1}{3}}(x)-W_{\rm \frac{1}{2},\frac{5}{6}}(x)W_{\rm -\frac{1}{2},\frac{5}{6}}(x)$. Clearly, both the emissivity and absorption coefficient depend on the comoving magnetic field strength and nonthermal particle distribution. Assuming that a constant portion of the dissipation power goes into the nonthermal electrons, we can find the nonthermal electron energy density $u'_{\rm e}$ and the electromagnetic energy density $u'_{\rm B}$ in the comoving frame,
\begin{equation}
\begin{aligned}
u'_{\rm e} &=\int_{ 1}^{\infty} d\gamma'~n'(\gamma')\gamma' m_{\rm e}c^2= \frac{\eta P_{\rm diss}}{\pi c (\theta_{\rm j} z \Gamma)^2}=\frac{4\eta \epsilon^2 L_{\rm j}(1-\chi)^k}{\pi c \theta^2_{\rm j} l^2_{\rm min}\Gamma^6_{\rm \infty}\zeta \chi^4} \\
u'_{\rm B} &=\frac{B^{\prime\,2}}{4\pi}=\frac{L_{\rm p}}{\pi c (\theta_{\rm j} z \Gamma)^2}=\frac{4\epsilon^2L_{\rm j}(1-\chi)}{\pi c\theta^2_{\rm j} l^2_{\rm min} \Gamma^6_{\rm \infty}\zeta^2\chi^2} ~~.
\end{aligned}
\label{eq:ueub}
\end{equation}
Generally speaking, the large-scale radio emission of blazars does not exhibit fast variability, indicating that the underlying nonthermal particles are in a quasi-stationary state. For simplicity, we assume that the nonthermal particles are of a straight power-law distribution,
\begin{equation}
n'(\gamma')=n'_{\rm 0}\gamma^{\prime\,-p}~~.
\end{equation}
In this way, both the emissivity and the absorption coefficient can be expressed as functions of the jet distance $z$ from the central engine $j'_{\rm \nu'}(z)$ and $\alpha'_{\rm \nu'}(z)$.

Based on Section \ref{sec:acceleration} and \citet{Giannios2019}, the jet acceleration is gradual over many orders of jet length scales from the central engine. Therefore, we neglect any bulk acceleration effects in the radiative transfer calculation. The intensity in the comoving frame is then given by
\begin{equation}
I'_{\rm \nu'}=\int_{ 0}^{\tau'_{\rm \nu'}}d\tau''_{\rm \nu'}~\frac{j'_{\rm \nu'}}{\alpha'_{\rm \nu'}}e^{-(\tau'_{\rm \nu'}-\tau''_{\rm \nu'})}~~,
\end{equation}
where $\tau'_{\rm \nu'}$ is the optical depth in the comoving frame given by
\begin{equation}
\tau'_{\rm \nu'}=\int_{ 0}^{s'}ds'~\alpha'_{\rm \nu'}~~,
\end{equation}
and $s'$ is the distance that a light beam travels along the line of sight. The intensity in the observer's frame is Doppler boosted as $I_{\rm \nu}=\delta^3I'_{\rm \nu'}$, where $\nu=\delta\nu'$.

With the above treatment, the large-scale jet emission is governed by four key parameters from the striped jet model, namely, the jet power $L_{\rm j}$, the stripe distribution profile parameters $l_{\rm min}$ and $a$, as well as the nonthermal particle spectral index $p$. In the following, we examine the dependence of observables on these key parameters. We choose the viewing angle for blazars at $\sin \theta_{\rm obs}=\Gamma^{-1}_{\rm \infty}$. Other parameters have trivial effects on the jet physical quantities and radiation signatures. For instance, the jet opening angle is typically a couple of degrees; within this range, the radiation signatures appear nearly the same.

\subsection{Core-Shift Effect}
\label{sec:coreshift}

\begin{figure}
\centering
\includegraphics[width=\linewidth]{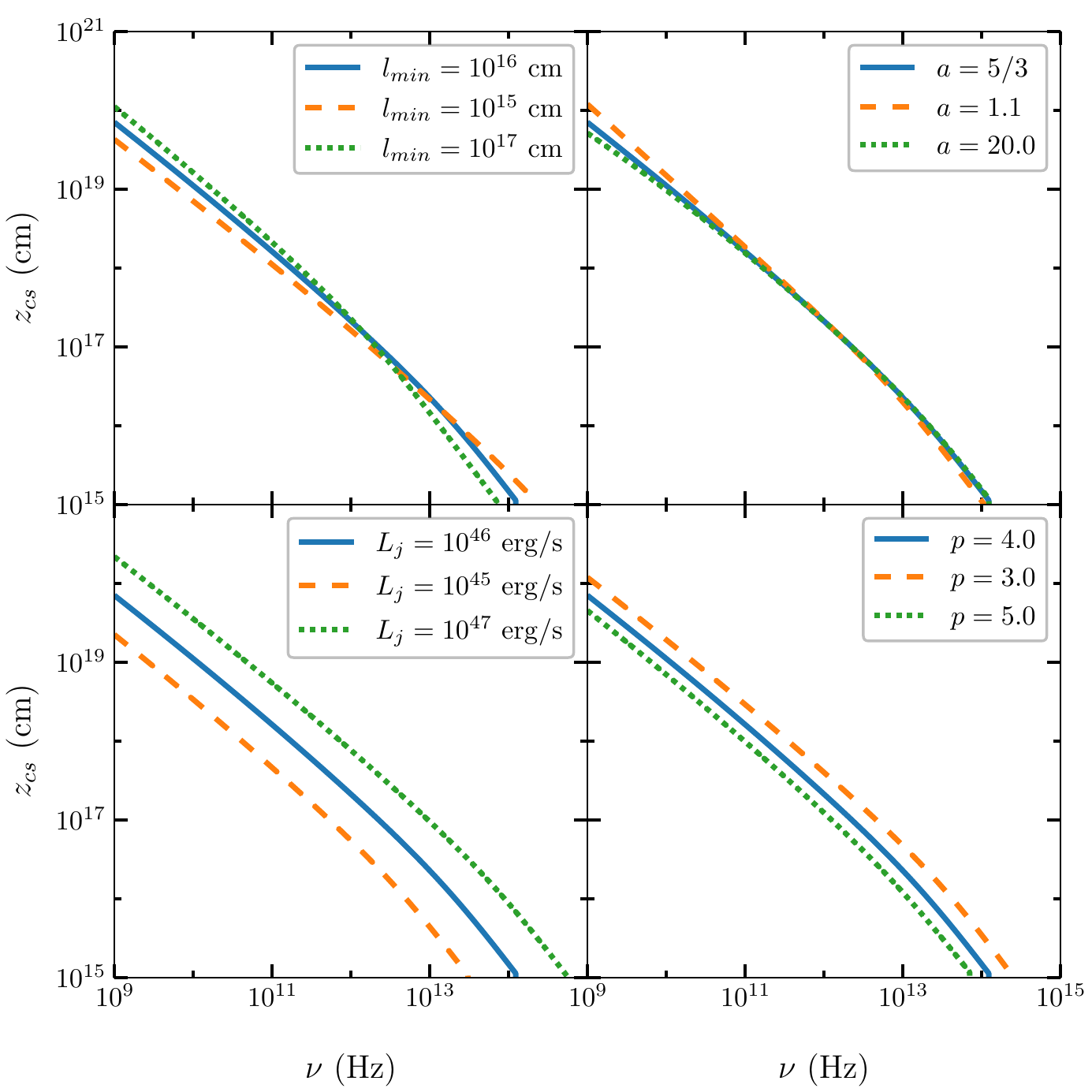}
\caption{The core-shift distance from the central engine at various observational frequencies. The four panels show the effects of four key parameters of the striped jet model, namely, the minimal stripe width $l_{\rm min}$ (upper left), the stripe profile index $a$ (upper right), the total jet power $L_{\rm j}$ (lower left), and the particle spectral index $p$ (lower right). The blue solid curve represents the default model parameters, while orange dashed and green dotted curves correspond to two variants of the parameters.}
\label{fig:coreshift}
\end{figure}

As shown in Equation \ref{eq:ssa}, the synchrotron self absorption coefficient depends on the nonthermal particle density, magnetic field strength, and observational frequency. \citet{Blandford1979} have shown that for a conical jet with decreasing nonthermal particle density and magnetic field from the central engine, the apparent size of the radio core, which is the distance where the optical depth $\tau_{\rm \nu}=1$, drops with higher observational frequency $\nu$. Thus, this interpretation needs a physical model for the postulated nonthermal particle and magnetic field profiles. Here we demonstrate how the core-shift relation can be explained under the striped jet model.

To analytically characterize the core-shift dependence on the striped jet parameters, we first recognize that the viewing angle in the comoving frame is given by
\begin{equation}
\sin \theta'_{\rm obs}=\delta \sin \theta_{\rm obs}~~.
\end{equation}
A light ray, emitted from the central spine of the jet, travels a distance of $s'=r/(\sin \theta_{\rm obs}\delta)$ in the comoving frame before it reaches the edge of the jet, where $r=z\tan\theta_{\rm j}$ is the cross-sectional radius of the jet at distance $z$. We can thus find the jet length that is covered in the light trajectory $s'$ as
\begin{equation}
\Delta z=\frac{r(\cos{\theta_{\rm obs}}-\beta)}{\sin{\theta_{\rm obs}}}~~.
\end{equation}
For typical blazar parameters, $\sin\theta_{\rm obs} \sim \Gamma^{-1}_{\rm \infty}$ and $\theta_{\rm j}$ is very small, we recognize that $\Delta z \ll z$ at any specific $z$. Since the jet bulk Lorentz factor $\Gamma$, the comoving nonthermal particle density $n'(\gamma')$, and the comoving magnetic field strength $B'$ all depend on $z$, within the light trajectory across the jet, they can be all be approximated as constants. Therefore, we obtain
\begin{equation}
\tau'_{\rm \nu'}=\alpha'_{\rm \nu'} \frac{\theta_{\rm j} z}{\sin{\theta_{\rm obs}} \delta} ~~.
\label{eq:opticaldepth}
\end{equation}
The specific synchrotron power $P'_{\rm \nu'}(\gamma')$ expression Equation \ref{eq:synpower} is very complicated to solve analytically. Here we take the $\delta$-function approximation \citep{Boettcher2012},
\begin{equation}
P_{\rm \nu'}^{\prime\,\delta}(\gamma')=\frac{4}{3}c\sigma_{\rm T}u'_{\rm B}\gamma^{\prime\,2}\delta(\nu'-\nu'_{\rm c}) ~~.
\end{equation}
Put this into Equations \ref{eq:ssa} and \ref{eq:opticaldepth}, we find the dependence of $\tau'_{\rm \nu}$ ($\nu=\delta \nu'$) on striped jet parameters as
\begin{equation}
\tau'_{\rm \nu} \propto \nu^{-\frac{p+4}{2}} (1-\chi)^{\frac{p+4k+2}{4}} \zeta^{-\frac{p+2}{2}} \chi^{-4} ~~.
\end{equation}
The above equation cannot be solved analytically, but from the asymptotic expressions (Equation \ref{eq:asymptotic}) and by setting $\tau'_{\rm \nu'}=1$, we can find the asymptotic expressions for high observational frequencies where the core-shift distance $z_{\rm cs}$ is very small,
\begin{equation}
z_{\rm cs} \propto \nu^{-\frac{3p+12}{3p+14}} ~~,
\end{equation}
and for low observational frequencies where the core-shift distance $z_{\rm cs}$ is very large,
\begin{equation}
z_{\rm cs} \propto \nu^{-\frac{2kp+8k-2p-8}{2kp+8k-p-2}} ~~.
\end{equation}
Typically, the nonthermal electrons that are responsible for the large-scale radio jet emission have a spectral index $p$ in the range from $\sim 3$ to $\sim 5$, and for the stripe distribution profile index $a>1$, we have $k>1.5$. With these parameters, one can easily see that the indices in both asymptotic expressions yield approximately $-1$. Therefore, the striped jet model predicts that the size of the unresolved radio core is anti-proportional to the observational frequency.

Figure \ref{fig:coreshift} presents the core-shift distance at different observational frequencies via the detailed radiative transfer outlined in Section \ref{sec:radiation}. The blue curve is the same default case in all panels, with $l_{\rm min}=10^{16}~\rm{cm}$, $a=5/3$, $L_{\rm j}=10^{46}~\rm{erg/s}$, and $p=4$. It is clear that the core-shift distance is generally anti-proportional to the observational frequency, with rather straight profile across several orders of observational frequencies. The profile becomes slightly curved when the core-shift surface is close to the central engine. This feature is not a robust prediction of the model because the expressions used in the previous analytical solutions fail to describe the jet acceleration and collimation close to the source. In reality, both $\Gamma$ and $z$ are nontrivial near the jet base, introducing boundary corrections there. This leads to the slightly curved core-shift profile at high frequencies. We also observe that the core-shift distance depends on the jet power $L_{\rm j}$ and the electron spectral index $p$. This is easy to understand, since the synchrotron self absorption is stronger with higher particle density, and both small $p$ and large $L_{\rm j}$ can increase the particle density.

The jet becomes completely optically thin for high-frequency bands for which $z_{\rm cs}$ reaches the jet base. At these frequencies, the synchrotron spectrum is described by the well-known power-law index $-(p-1)/2$. We can find this spectral break frequency analytically using the asymptotic expressions,
\begin{equation}
\nu_{\rm b} \propto L_{\rm j}^{\frac{p+6}{2p+8}}l_{\rm min}^{\frac{2}{3p+12}}~~.
\label{eq:nub}
\end{equation}
One can see that $\nu_{\rm b}$ is fairly insensitive to the minimal stripe width $l_{\rm min}$ and does not depend on the stripe profile index $a$. Rather, it depends on the jet power $L_{\rm j}$ with an index of $\sim 0.5$. These trends are consistent with our numerical findings in Figure \ref{fig:coreshift}.

\subsection{Overall Radio Spectrum}
\label{sec:fullradiospec}

\begin{figure}
\centering
\includegraphics[width=\linewidth]{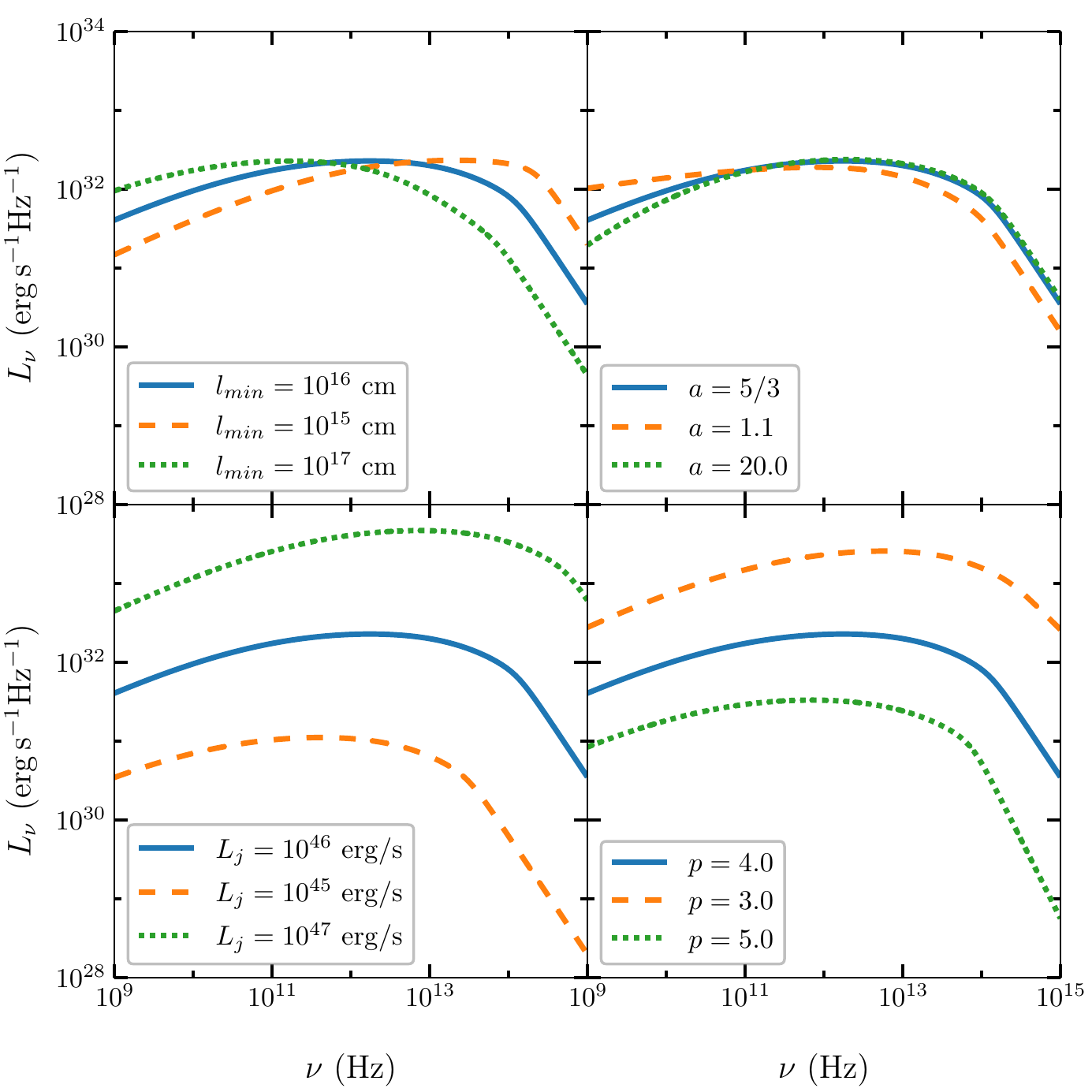}
\caption{The synchrotron spectrum $L_{\rm \nu}$ for the large-scale striped jet blazar emission. The emission spectrum is fairly flat, i.e., $L_{\rm \nu}\sim \nu^0$, over a wide range of frequencies. Panels and curves are consistent with Figure \ref{fig:coreshift}.}
\label{fig:radiospec}
\end{figure}

Using decreasing nonthermal particle and magnetic field profiles, \citet{Blandford1979} have shown that the synchrotron emission and self absorption result in the observed flat-to-inverted radio spectrum. This suggests that large-scale jet spectral properties can be powerful diagnostics of the jet energy dissipation profile as a function of distance from the source. Here we illustrate that the large-scale jet spectrum can be explained under the striped jet model and constrain key model parameters on jet launching and dissipation.

The complete radiative transfer is too complicated to solve analytically. Here, we make a number of simplifying assumptions in order to derive approximate, but illuminating, analytic expressions for the jet emission and then we proceed with the presentation of the fully numerical results. So as to characterize the large-scale jet spectral properties, we notice that the jet can be treated as optically thin at a given observational band $\nu$ for jet length $z>z_{\rm cs}(\nu)$. Given our  choice for the viewing angle of $\sin \theta_{\rm obs}=\Gamma^{-1}_{\rm \infty}$, the comoving viewing angle is not perpendicular to the jet propagation direction except for the asymptotic jet scale, where the bulk Lorentz factor is $\Gamma_{\rm \infty}$. However, the resulting integration for the jet luminosity is still very complicated, requiring further simplifications. Aforementioned, the light ray trajectory is almost in the radial direction at any given $z$, because $\Delta z \ll z$. Therefore, we simply take the comoving viewing angle to be in the radial direction for all $z$, and integrate the emission over the entire jet from $z_{\rm cs}$, i.e.,
\begin{equation}
L_{\rm \nu}=\int_{ z_{\rm cs}}^{\infty} dz~\delta^3\pi r^2 j'_{\rm \nu'} ~~,
\end{equation}
where $r=z\tan \theta_{\rm j}$ is the jet radius at $z$. Putting the $\delta$-function approximation for the specific synchrotron power $P'_{\rm \nu'}(\gamma')$ into the above equation, we find
\begin{equation}
L_{\rm \nu} \propto \nu^{-\frac{p-1}{2}} \int_{ \zeta_{\rm cs}}^{\infty} d\zeta~ (1-\chi)^{\frac{p+4k+1}{4}} \zeta^{-\frac{p-1}{2}} \chi^{-2} ~~,
\label{eq:radiospec}
\end{equation}
where $\zeta_{\rm cs}=(2\epsilon z_{\rm cs})/(l_{\rm min}\Gamma^2_{\rm \infty})$. Similar to the treatment in Section \ref{sec:coreshift}, we take the asymptotic expressions for low observational frequencies ($\chi \rightarrow 1$),
\begin{equation}
L_{\rm \nu} \propto \nu^{\frac{2p+13}{2kp+8k-p-2}}~~,
\end{equation}
and for high observational frequencies ($\chi \ll 1$),
\begin{equation}
L_{\rm \nu} \propto \nu^{-\frac{2p+3}{3p+14}}~~.
\end{equation}
For the typical nonthermal particle spectral index $p$ and the stripe profile index $a$, we can see that the radio spectral index $\alpha$ ($L_{\rm \nu}\propto \nu^{\alpha}$) lies in the range of $\sim 0.5$ to $\sim -0.5$ from low to high frequencies. Thus the spectrum first rises slowly at low frequencies and then drops at high frequencies.

However, the above calculations have one caveat: the asymptotic expressions are not obtainable in practice. As shown in the previous section, for typical blazar parameters, the core-shift distance hardly reaches adequately large number so that $\chi \rightarrow 1$; on the other hand, even at the base of the jet $\Gamma>1$, so that $\chi \ll 1$ does not apply. As shown in the following numerical calculations, the observed spectral shape is much closer to a flat or slightly inverted spectrum, with spectral index $\alpha$ between $0.2$ to $0$. Nonetheless, one spectral feature remains: the overall radio spectrum is not exactly straight, but appears slightly curved with a peak at $\nu_{\rm peak}$. Since $z_{\rm cs} \propto \nu^{-1}$ and by observing Equation \ref{eq:radiospec}, we can see that the spectrum becomes flat if the integration is proportional to $\zeta_{\rm cs}^{-(p-1)/2}$. This implies that the emission at $\nu_{\rm peak}$ is dominated by the jet segment at $z \propto l_{\rm min}\Gamma^2_{\rm \infty}$. We find that the above expression is proportional to the peak of the dissipation profile $z_{\rm peak}$, thus $\nu_{\rm peak} \propto z_{\rm peak}^{-1}$. Therefore, for a given $\Gamma_{\rm \infty}$, the peak of the radio spectrum is anti-proportional to the minimal stripe width $l_{\rm min}$.

Figure \ref{fig:radiospec} presents the large-scale synchrotron spectrum based on the numerical solution of the full radiative transfer equations in Section \ref{sec:radiation}. The jet luminosity naturally depends on the jet power $L_{\rm j}$ and the electron spectral index $p$, as shown in the lower two panels. The stripe profile index $a$ also affects the spectral shape of the partially optically thin part of the spectrum. \citet{Giannios2019} have shown that $a$ governs the jet dissipation profile, where smaller $a$ implies more energy dissipated at larger jet distances from the central engine. As shown in Section \ref{sec:coreshift}, the low-frequency bands are less absorbed at large jet distance. Consequently, the low-frequency flux slightly rises with smaller $a$, leading to a flatter spectrum. Obviously, $p$ also affects the synchrotron spectral shape. Ideally its value can be directly constrained with the fully optically thin part of the synchrotron spectrum. As a result, although the effect is very small, the radio spectral shape can shed light on the stripe distribution profile.

The spectrum appears soft for high-frequency bands where the jet becomes fully optically thin. We can see in Figure \ref{fig:radiospec} that the dependence of this frequency on striped jet parameters fits well with Equation \ref{eq:nub}. We remind the readers that, under the striped jet model, the frequency $\nu_{\rm b}$ that the jet becomes fully optically thin is not the same as the peak of the large-scale spectrum at $\nu_{\rm peak}$. As shown in Equation \ref{eq:nub}, $\nu_{\rm b}$ has trivial dependence on $l_{\rm min}$, but $\nu_{\rm peak} \propto l_{\rm min}^{-1}$. This is clearly shown in the upper left panel of Figure \ref{fig:radiospec}. In this parameter study we choose three different values for $l_{\rm min}$ separated by two orders of magnitude. We can see that the spectral peaks $\nu_{\rm peak}$ are separated by one order from each other, but the spectral break frequencies $\nu_{\rm b}$ are comparable. Therefore, the separation of $\nu_{\rm peak}$ and $\nu_{\rm b}$ implies the minimal stripe width $l_{\rm min}$, which characterizes the quasi-periodic magnetic field reversal produced by the MRI dynamo at the central engine.

\subsection{Radio Spectral Evolution with Jet Propagation}
\label{sec:radiospecvsz}

\begin{figure}
\centering
\includegraphics[width=\linewidth]{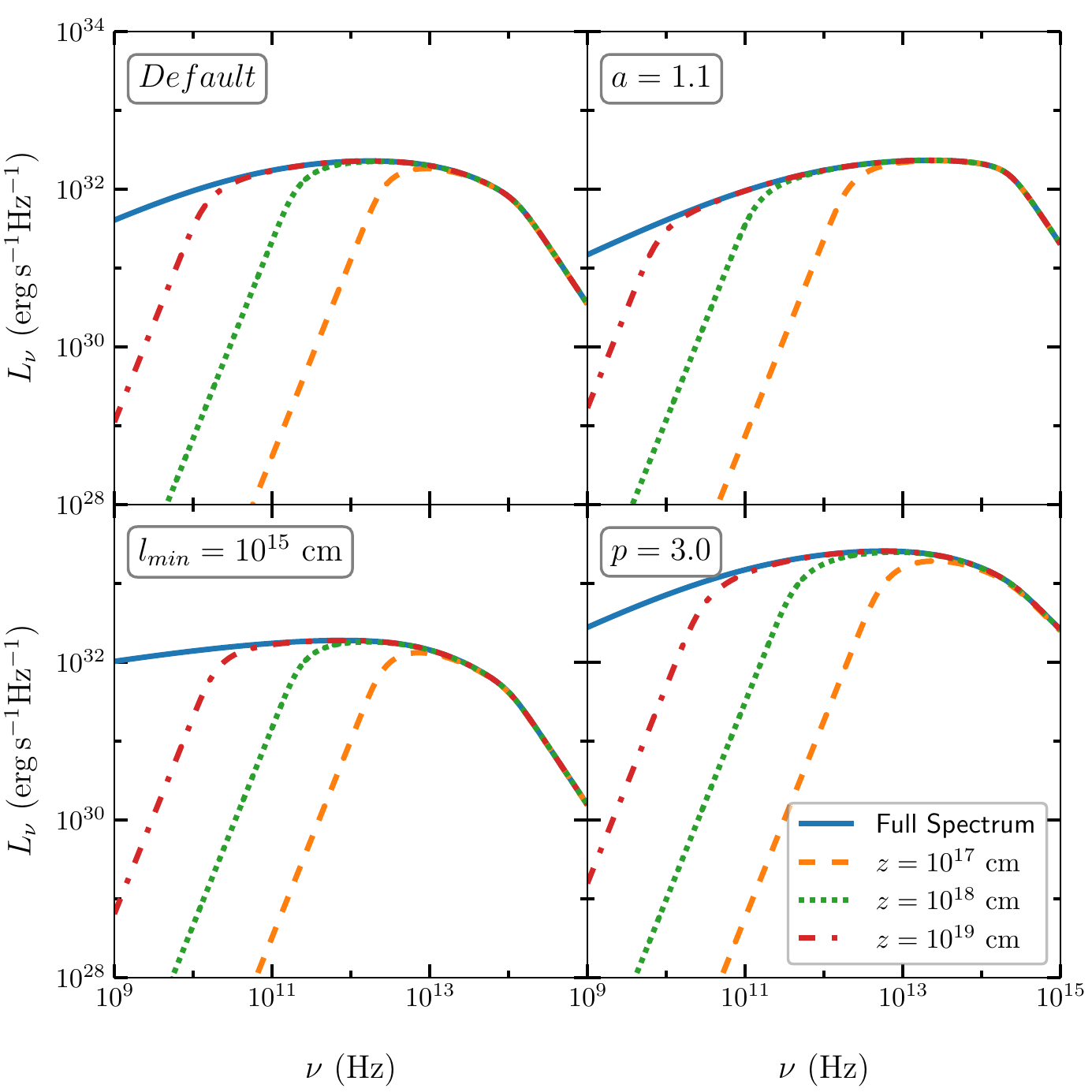}
\caption{The synchrotron spectrum $L_{\rm \nu}$ integrated to different jet distances from the central engine. The four panels are the default parameters (upper left), a smaller minimal striped width $l_{\rm min}=10^{15}~\rm{cm}$ (upper right), a smaller stripe profile index $a=1.1$ (lower left), and a smaller electron spectral index $p=3.0$ (lower right). For the last three panels, except for the said changes, all other parameters are the same with the default case.}
\label{fig:radiospecvsz}
\end{figure}

VLBI techniques can resolve the jet structure and probe the synchrotron spectral properties at different distances from the central engine. We find that the striped jet model predicts a more inverted spectral shape at smaller distances from the central engine. Therefore, the radio core spectrum should appear harder than the overall spectrum under the striped jet model.

Figure \ref{fig:radiospecvsz} shows the radio spectral evolution along the jet propagation direction. This is done by integrating the synchrotron intensity from the central engine to the given jet distances. In addition to the default parameters, we plot one sample each for different $l_{\rm min}$, $a$, and $p$. This is because Figure \ref{fig:radiospec} shows that $L_{\rm j}$ has no effect on the spectral index. It is clear that at smaller jet distances from the central engine, the low-frequency spectra are significantly absorbed, so that only the high-frequency emission can be observed, consistent with the core-shift relation. For the partially optically thin part of the spectrum, it appears more inverted at smaller jet distances, with the spectral index reaching as hard as $\alpha \sim 0.5$. This suggests that the radio core spectrum should be harder than the overall radio spectrum. Figure \ref{fig:radiospecvsz} suggests that this feature holds for any striped jet model parameters.

Additionally, we notice that the spectra appear the same at various distances for high-frequency bands ($\nu\gtrsim \nu_{\rm b}$). Comparing with Fig. \ref{fig:evolution}, one can see that $z=10^{17}~\rm{cm}$ locates before the dissipation peak $z_{\rm peak}$. This suggests that the high-frequency emission is dominated by the contribution near the jet base instead of the peak of dissipation. The reason is that the synchrotron self absorption is very small for high-frequency bands, thus most emission comes from the jet base, where the magnetic field is much stronger. A very interesting consequence is that the emission at the spectral luminosity peak $\nu_{\rm peak}$ is dominated by the emission from the jet segment a small factor larger than $z_{\rm cs}(\nu_{\rm peak})$. From Figure \ref{fig:radiospecvsz}, this location is at $\sim 10^{18}~\rm{cm}$. Meanwhile, as mentioned in Section \ref{sec:fullradiospec}, the luminosity at the spectral peak is dominated by the emission near the peak of the dissipation power $z_{\rm peak}$. Consider $z=\Gamma_{\infty}^2l_{min} \sim 300 l_{\rm min}=3\times 10^{18}~\rm{cm}$, which scales well with the dissipation peak $z_{\rm peak}$, we find that the above two estimates of the large-scale jet synchrotron emission peaks are consistent. Apparently, the size of the compact core at $\nu_{\rm peak}$ can be a good indicator of the peak of the dissipation power, which can help to locate the blazar zone and the minimal stripe width at the central engine.

\section{Blazar Zone}
\label{sec:blazarzone}

The highly variable, nonthermal-dominated multi-wavelength emission from the blazar jet suggests the presence of a very localized blazar zone, where a large amount of the jet energy is efficiently dissipated to accelerate nonthermal particles and radiate. Under the striped jet model, this blazar zone should locate within the broad peak of the dissipation profile. We find that it should lie between 0.1 to a few parsecs from the central engine, where the jet is moderately magnetized, with $\sigma \gtrsim 1$. The broadband emission and flares are driven by relativistic magnetic reconnection between stripes. The magnetic field strength in the blazar zone is typically between $\sim 0.1~\rm{G}$ to $\sim 10~\rm{G}$. Under these physical conditions, we can generate the typical two-component blazar spectral energy distribution (SED). Importantly, the radio part of the SED, which is due to the large-scale jet emission and often ignored in the blazar SED studies, can be consistently produced under the striped jet model by combining large-scale synchrotron and the blazar zone emission. In this section, we first investigate the blazar zone location and physical conditions, then describe its radiation modeling, finally we study the multi-wavelength radiation signatures.

\subsection{Location and Physical Properties}
\label{sec:location}

\begin{figure}
\centering
\includegraphics[width=\linewidth]{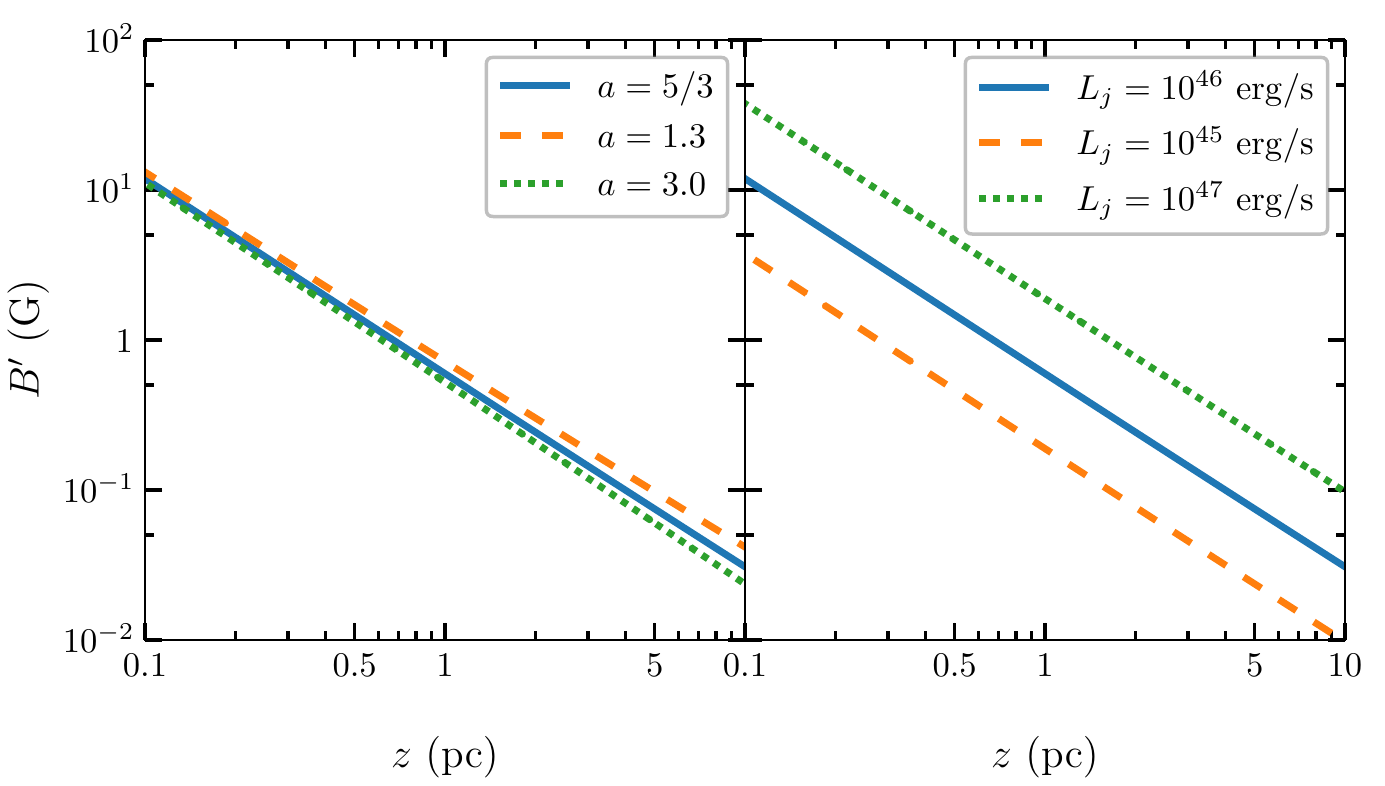}
\caption{The local comoving magnetic field strength $B'$ between 0.1 and 10 pc from the central engine for various stripe profile index $a$ (left panels) and total jet power $L_{\rm j}$ (right panels).}
\label{fig:magneticfield}
\end{figure}


The blazar zone is a unique region in the blazar jet. This region shows very bright multi-wavelength emission, implying strong dissipation of the jet energy. The emission is nonthermal-dominated, and spectral fitting models typically find that the blazar zone has a bulk Lorentz factor of $\gtrsim 10$ \citep{Boettcher2013,Paliya2018}. Additionally, the multi-wavelength emission is variable, in particular, infrared to optical emission often flares together with X-rays and $\gamma$-rays \citep{Chatterjee2012,Liodakis2019}. Given the typical day-to-week variability time scales, causality relation requires that the blazar zone must be very small, usually of sub-parsec size. Since the jet is highly magnetized at the launching site, the blazar zone may be the transition location where the jet energy changes from magnetic-dominated to kinetic-dominated condition. Therefore, it is crucial to study the blazar zone and the large-scale jet in a unified physical picture to probe the physical conditions of the blazar zone and the related large-scale jet evolution.

Under the striped jet model, we suggest that the blazar zone should match four criteria to match observational properties. First, it should be located within the broad peak of the jet dissipation profile (so that there is sufficient energy to power the observed blazar emission). Second, it has a bulk Lorentz factor $\Gamma \gtrsim 10$. Third, since the blazar flares are driven by efficient magnetic reconnection under the striped jet model, the blazar zone should be magnetized, with the magnetization factor $\sigma\gtrsim1$.\footnote{For $\sigma\lesssim 1$, the reconnection process turns slower and the particle acceleration spectra become too steep to account for observations \citep{Guo2015,Sironi2015}.} Finally, since many blazar flaring activities exhibit highly variable infrared to optical emission, the blazar zone should be optically thin to infrared and higher frequencies. 
We can see from Figure \ref{fig:evolution} that the broad dissipation peak, which scales as $z=\Gamma_{\infty}^2l_{min}$, matches well with the above four criteria. Therefore, the blazar zone should locate between 0.1 pc to a few pc. These blazar zone properties come naturally under the striped jet model, because the broad dissipation peak results from the most efficient magnetic reconnection between stripes, so that the blazar zone is still sufficiently magnetized but a large amount of the magnetic energy has already converted into bulk kinetic energy. As shown in Figure \ref{fig:coreshift}, within this distance range the jet is optically thin to the infrared and higher-energy bands. We find that our estimate of the blazar zone location is consistent with observations \citep{Aleksic2011,Hayashida2012}. Figure \ref{fig:evolution} shows that the presence of the blazar zone requires the minimal stripe width $l_{\rm min}\lesssim 10^{16}~\rm{cm}$ and the stripe index $a\gtrsim 1.3$, otherwise the bulk Lorentz factor is too small. Considering that the observed jet acceleration beyond the blazar zone requires $l_{\rm min}\gtrsim 10^{16}~\rm{cm}$ and $1.1\lesssim a\lesssim 3.0$, these two phenomena put strong constraints on the striped jet model parameters, $l_{\rm min}\sim 10^{16}~\rm{cm}$ and $1.3\lesssim a\lesssim 3.0$.

The comoving magnetic field strength at $\sim 1~\rm{pc}$ should be approximately 1 G under the striped jet model. From Equation \ref{eq:ueub}, we can easily see that
\begin{equation}
B' \propto \sqrt{\frac{L_{\rm j}(1-\chi)}{z^2 \chi^2}}~~,
\end{equation}
in which $B'$ is proportional to $\sqrt{L_{\rm j}}$ and $z^{-1}$. In principle, $B'$ also implicitly depends on $l_{\rm min}$ and $a$ through the $\chi$ terms, but $(1-\chi)/\chi^2$ is typically on the order of 1 for most part of the jet. From Equation \ref{eq:ueub}, we can find that the magnetic field strength close to the central engine ($z=10^{15}~{\rm cm}$) is
\begin{equation}
B'\sim 2000~\rm{G}~ \Big(\frac{L_{\rm j}}{10^{46}~\rm{erg/s}}\Big)^{\frac{1}{2}}~~,
\end{equation}
where $L_{\rm j}=10^{46}~\rm{erg/s}$ is the Eddington luminosity for a supermassive black hole of $10^8$ solar mass. Therefore, the comoving magnetic field strength in the blazar zone at $\sim 1~\rm{pc}$ is about $0.7~\rm{G}$, consistent with most leptonic blazar models \citep{Boettcher2013}.

Figure \ref{fig:magneticfield} shows the comoving magnetic field strength between 0.1 pc to 10 pc based on numerical calculations. The results are consistent with the above analytical expressions, and we can see that the magnetic field strength covers a relatively large range from $\sim 0.01~\rm{G}$ to $\sim 10~\rm{G}$. Meanwhile, Figure \ref{fig:evolution} shows that the dissipation power profile in this distance range is nearly flat, at $P_{\rm diss}\sim 10^{45}~\rm{erg/s}$. These parameters are in good agreement with leptonic blazar models. Spectral fitting models often find that the magnetic field strength is relatively small ($B'\lesssim 0.1~\rm{G}$) for BL Lacs, but higher $B'~\sim 0.5~\rm{G}$ for flat spectrum radio quasars. Under the striped jet model, this implies that for flat spectrum radio quasars, either the blazar zone is closer to the central engine, or the jet intrinsically has higher power. In both cases, the external photon field in the blazar zone is higher, naturally explaining the strong external Compton contribution often seen in flat spectrum radio quasars. For the hadronic models, however, the striped jet model requires a very high total jet power, at $L_{\rm j}\sim 10^{48}~\rm{erg/s}$, which is about ten times of the Eddington luminosity for $10^9$ solar mass supermassive black holes. Under this extreme condition the blazar zone can have magnetic field strength at $\sim 100~\rm{G}$ and dissipation power at $P_{\rm diss}\sim 10^{47}~\rm{erg/s}$, consistent with hadronic spectral fittings.

\subsection{Blazar Model}
\label{sec:blazarmodel}

The blazar zone emission exhibits two spectral components. The low-energy component extending from radio to optical, and in some cases up to soft X-rays, is dominated by synchrotron emission by nonthermal electrons. The high-energy component covering X-rays and $\gamma$-rays is often interpreted as the inverse Compton scattering of low-energy photons by the same electrons \citep{Dermer1992,Sikora1994}. Alternatively, the high-energy spectral component may be of hadronic origin, whose emission comes from the proton synchrotron and hadronic cascades \citep{Mannheim1993,Muecke2003}. Both low- and high-energy spectral components are variable, which indicates efficient particle acceleration and cooling. Additionally, the particles responsible for blazar flares must be quickly accelerated in situ, so that the nonthermal particle distributions are likely different from those of the large-scale jet.

Under the striped jet model, the nonthermal electrons are accelerated via relativistic magnetic reconnection. This non-ideal plasma physical process can rearrange magnetic field lines of opposite polarity, dissipating a large amount of magnetic energy to accelerate particles during this process. Numerical simulations of relativistic magnetic reconnection have demonstrated the formation of an extended power-law in the particle energy distribution in both pair and electron-proton plasma \citep{Guo2014,Guo2016,Sironi2014}. Relativistic magnetic reconnection is efficient if the blazar zone is magnetized, i.e., $\sigma>1$. We also need to include the adiabatic and radiative cooling for the nonthermal particles. In particular, the synchrotron and Compton scattering cooling effects are significant for nonthermal electrons. To include all the above effects, we employ the one-zone blazar radiation code developed by \citet{Boettcher2013}. Given the magnetic field and nonthermal particle injection, this code can self-consistently evaluate the radiative and adiabatic cooling in the blazar emission region.




\subsection{Multi-Wavelength Radiation Signatures}
\label{sec:blazarrad}


\begin{figure}
\centering
\includegraphics[width=\linewidth]{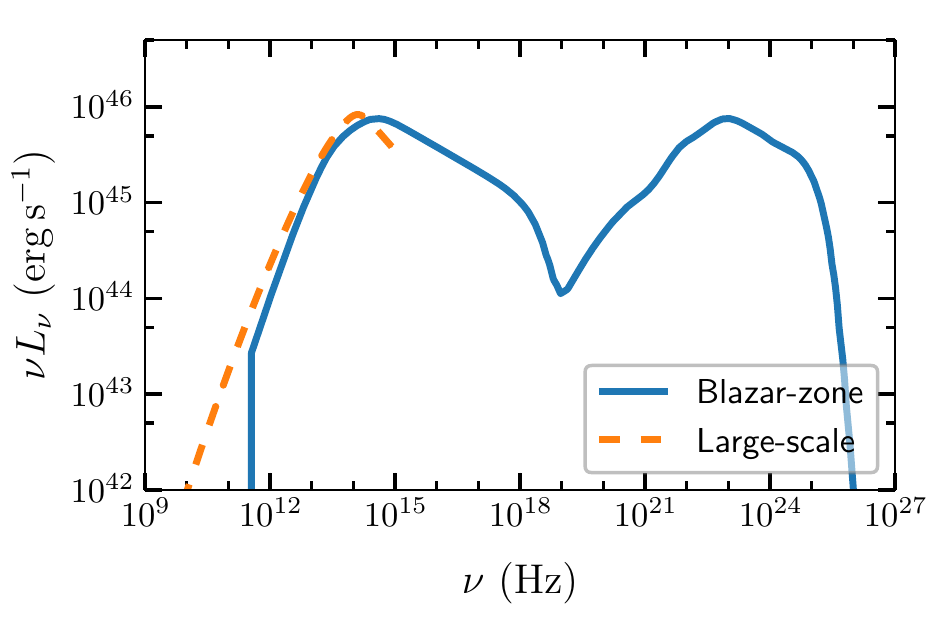}
\caption{A sample of the blazar SED under the striped jet model. The blue solid line represents emission from the flaring region, while the orange dashed line represents the quasi-stationary large-scale jet emission.}
\label{fig:blazarsed}
\end{figure}

Figure \ref{fig:blazarsed} shows the blazar SED in the quasi-stationary state. We assume that the emission region is at $1~\rm{pc}$ from the central engine, and the supermassive black hole is $10^8$ solar mass accreting at the Eddington limit. The thermal emission from the accretion disc gets reprocessed by the broad line region and dusty torus, which acts as the external photon field for the Compton scattering. The magnetic field strength and nonthermal particle energy density are consistently derived from the striped jet model using the default parameters. Given that the nonthermal electrons in the blazar zone are freshly accelerated by very efficient magnetic reconnection, we take a harder power-law spectrum with $p=2.5$ and $\gamma_{\rm min}=2000$. We can see that the above parameters can generate typical blazar SED. Importantly, the large-scale jet spectrum overlaps with the blazar zone spectrum very well at infrared to optical bands, then it quickly cuts off due to its soft particle spectral index. This suggests that the infrared to optical blazar emission has a significant contribution from the large-scale jet. At lower energies, the large-scale jet emission dominates over that from the blazar zone.

The above SED is given under the leptonic model. Generally speaking, the striped jet model does not predict considerable hadronic signatures, due to insufficient energy budget. As we can see in Figure \ref{fig:evolution}, the dissipation covers a broad range of distances, and the peak of the dissipation power is $\sim 10\%$ of the total jet power. For typical hadronic blazar models, the proton kinetic power is required to be $>10^{47}~\rm{erg/s}$ \citep{Boettcher2013,Paliya2018}. Therefore, although protons can still accelerate via reconnection as shown in recent numerical simulations \citep{Guo2016,Nishikawa2020,Petropoulou2018}, their contributions to the high-energy spectral component as well as neutrino emission are trivial under the striped jet model. However, if the central engine accretion is significantly super-Eddington, so that the total jet power can reach as high as $\gtrsim 10^{48}~\rm{erg/s}$, the striped jet model can exhibit significant hadronic signatures. In this situation, the magnetic field strength can reach as high as $\sim 100~\rm{G}$ at the peak dissipation scale suppressing the Compton scattering by nonthermal electrons and implying that the blazar high-energy spectral component may be dominated by proton synchrotron and hadronic cascades.

For the variability time scales, the striped jet model generally predicts a power spectral density profile with an index of $\sim 2.0$. This is because blazar flares are driven by magnetic reconnection between stripes. Recent numerical simulations have demonstrated that flares driven by reconnection result from the plasmoid production and mergers in the current sheet \citep{Zhang2018,Zhang2020,Christie2019}. Since the electron acceleration and cooling are relatively fast in the blazar zone, the light crossing time scales of the plasmoids and stripes should govern the observed variability time scales. \citet{Petropoulou2018} have shown that the plasmoid extent follows a power-law distribution in the current sheet, up to a fraction of the current sheet size. The power-law index is typically around $2.0$, but can reach as low as $1.3$. Long-term variability, on the other hand, is related to the size of stripes. As shown above, for typical blazars the striped spectral index should be between $1.3-3.0$. Therefore, generally one does not expect any major breaks in the power spectral density from short- to long-term variability based on the striped jet model. In the case that there is a small break between short-term to long-term variability (for instance, if $a>2$), the break should happen at a time scale related to $l_{\rm min}$, which is likely on the order of one week in the observer's frame.

\section{Summary and Discussion}
\label{sec:discussion}

\begin{figure}
\centering
\includegraphics[width=\linewidth]{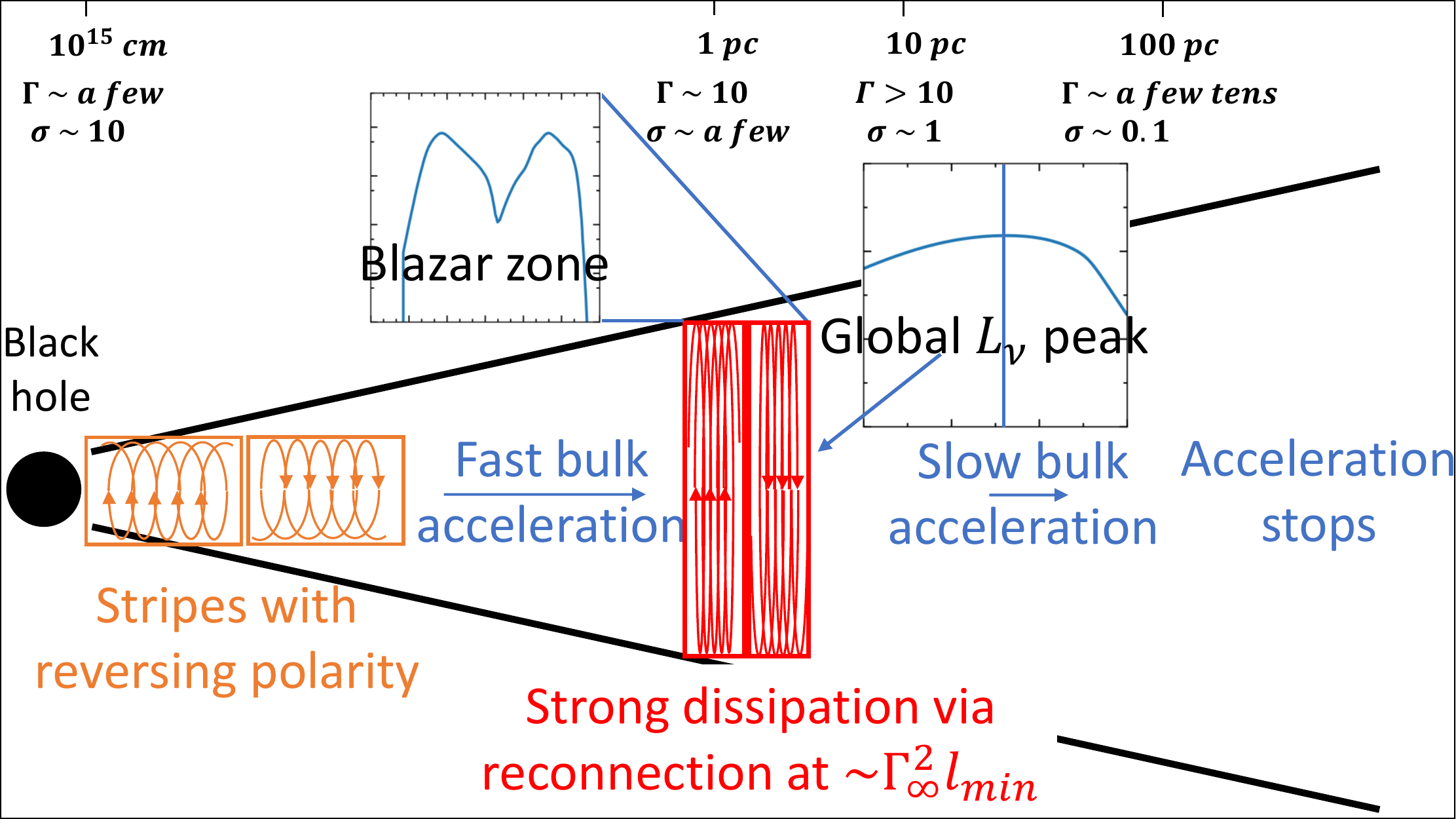}
\caption{A sketch of the striped jet model. The central engine generates stripes of toroidal-dominating magnetic fields with reversing polarity. The jet bulk acceleration is powered by magnetic reconnection between stripes. As the jet expands along its propagation, the stripes are stretched sideways, making them vulnerable to tearing instabilities, which trigger strong magnetic reconnection. This is the location of the blazar zone and contributes the most to the spectral luminosity peak of the large-scale emission. The blazar zone has a bulk Lorentz factor of $\Gamma \sim 10$ and magnetization factor of $\sigma>1$. The jet bulk acceleration is relatively fast when its magnetization is above 1. The jet magnetization drops below 1 at $\sim 10~\rm{pc}$, and the bulk acceleration slows down. At $\sim 100~\rm{pc}$, the jet becomes kinetic-energy-dominated, and the bulk acceleration stops.}
\label{fig:stripedjet}
\end{figure}

It is widely believed that the jet is highly magnetized at the launching site. However, the jet acceleration and energy dissipation during its propagation are not well understood \citep{Achterberg2001,Nishikawa2003,ONeill2012,Barniol2017,Dong2020}. In particular, the physical conditions at the blazar zone, where the jet dissipates a large amount of energy and emits strong multi-wavelength emission, are the key to the overall jet dynamics. Models diverge here into two general scenarios. One is that the jet has already converted most of its initial magnetic energy into bulk kinetic energy, so that blazar flares are driven by dissipation of the bulk kinetic energy, typically via shocks \citep{Marscher1985,Boettcher2010}. The other is that a large amount of the magnetic energy is dissipated at the blazar zone, so that the jet remains magnetized to at least about one parsec from the central engine \citep{Giannios2006,Giannios2009,Petropoulou2016,Zhang2017}. The striped jet model belongs to the latter scenario. Figure \ref{fig:stripedjet} illustrates the basic properties and features of the striped jet model. The striped jet is characterized by stripes of toroidal-dominating magnetic field with reversing polarity generated from the central black hole accretion. These stripes can undergo magnetic reconnection due to the anti-polarity toroidal magnetic fields. Initially the reconnection is not very efficient, due to the fast jet expansion. However, with the jet reaching larger distances of approximately one parsec, the stripes are stretched sideways, becoming unstable to tearing instabilities and the magnetic reconnection timescale becomes comparable to the jet expansion time, which results in very efficient magnetic reconnection. As a result, the striped blazar jet is moderately strongly magnetized at the blazar zone, and dissipates a large portion of its magnetic energy there. After that the reconnection efficiency drops, and the jet energy gradually transits from magnetic-dominated to kinetic-dominated. Still, the reconnection continues to convert magnetic energy to kinetic energy even at tens of parsecs, leading to further acceleration of the jet, which is consistent with recent VLBI observations \citep{Homan2015}. The observed acceleration at tens of parsecs from the central engine implies relatively hard power-law spectrum of the stripe distribution ($a<3.0$) and a minimal stripe width that is not too small ($l_{\rm min}\gtrsim 10^{16}$~cm).

The striped jet model can naturally explain the core-shift and flat-to-inverted large-scale jet spectrum, due to the continuous dissipation and nonthermal particle acceleration over a broad range of distances. The size of the compact radio core generally follows $z_{\rm cs}(\nu) \propto \nu^{-1}$ for any striped jet model parameters. Very interestingly, the spectral luminosity of the large-scale jet emission is slightly curved, and its peak is not due to that the jet becomes completely optically thin. Instead, the spectral peak marks the peak of the jet dissipation, and we find that it is essentially the size of the compact radio core at the spectral peak, $z_{\rm peak}\sim z_{\rm cs}(\nu_{\rm peak})$. Not only the dissipation power peak marks the location of the blazar zone, but it also points to a minimal stripe width at the central engine of $l_{\rm min}\sim 10^{16}$~cm. Therefore, the two observable properties of the large-scale jet, $\nu_{\rm peak}$ and $z_{\rm cs}(\nu_{\rm peak})$, can be very important diagnostics of the blazar zone location and black hole accretion.

The striped jet model suggests that the blazar zone is located within the broad peak of the jet dissipation profile, which is from 0.1 to a few parsecs. The local comoving magnetic field strength in the blazar zone scales as $B'\propto z^{-1}L_{\rm j}^{1/2}$, which lies in $0.01~\rm{G}$ to $10~\rm{G}$ for typical striped jet model parameters. The dissipation power at the blazar zone is approximately $10\% L_{\rm j}$. To generate typical blazar SEDs, it requires a relatively strong dissipation at the blazar zone, implying that the stripe power-law index to be $a>1.3$ and $l_{\rm min}\lesssim 10^{16}$~cm. The striped jet model can consistently produce the large-scale and blazar zone SEDs, where the radio part of the emission is dominated by the large-scale jet emission. The blazar zone emission takes over at approximately infrared bands, where the entire jet becomes optically thin. In the case that the jet power is super-Eddington, the striped jet model predicts a very high magnetic field strength at the blazar zone ($B'\sim 100~\rm{G}$), and the high dissipation power ($P_{\rm diss}\gtrsim 10^{47}~\rm{erg/s}$) can permit the hadronic high-energy spectral component via proton synchrotron and hadronic cascades. Since the blazar flares are driven by relativistic magnetic reconnection between stripes, the flaring time scales with the light crossing time of the plasmoids and stripe widths for short and long variability patterns, respectively \citep{Zhang2018,Christie2019,Zhang2020}. Since the plasmoid and stripe distribution profiles have similar power-law indices \citep{Petropoulou2018}, the striped jet model expects a straight power spectral density profile with an index between $\sim 1.3$ and $\sim 2.0$. Furthermore, optical polarization monitoring programs have suggested that blazars with large polarization angle rotations tend to be more active in both flux and polarization signatures \citep{Blinov2018}. Recent combined PIC and radiation transfer simulations have shown that such phenomena result from reconnection between nearly anti-parallel magnetic field lines in the blazar zone \citep{Zhang2020}. Therefore, blazars with large polarization angle rotation can be a signature of the striped jet morphology.

Taken all the constraints together, we infer from blazar observations that the minimum stripe width is $l_{\rm min}\sim 10^{16}$~cm, while the index that describes the stripe distribution is $1.3\lesssim a\lesssim 3$. The range of the index $a$ is consistent with the minimum stripes containing most of the power. Also, the value $a=5/3$ favored by \citet{Giannios2019} from general energy arguments falls within this range. The scale $l_{\rm min}$ may be arguably associated with the MRI growth timescale close to the inner edge of the disc; possibly at the location where the shear torques peak. For a zero-torque inner boundary disc model \citep{Novikov1973}, this location depends on the black-hole spin ranging from a few $R_{\rm g}$ (fast spin) to $\sim 20R_{\rm g}$ (no spin). For the sake of an estimate, we consider a black hole of $\sim 3\times 10^8 M_\odot$ and dimensionless spin parameter of $\sim 0.8$. The dissipation peak at the disc takes place at $\sim 8R_{\rm g}$, where the Keplerian period is $T_{\rm orb}\sim 20R_{\rm g}/c$. Consequently, the MRI growth timescale may be estimated to be $\sim 10 T_{\rm orb}$, resulting in stripes with $l_{\rm min}\sim 10cT_{\rm orb}\sim 10^{16}$~cm. Our favored values for $l_{\rm min}$ are, therefore, not unexpected.   

Magnetic reconnection in the stripes is not the only potential dissipative mechanism in a magnetically dominated jet. Kink-type instabilities are also likely to happen in a magnetized jet. Recent 3-dimensional MHD jet simulations have revealed that kink instabilities can twist the magnetic field and build up current sheets as well as drive strong turbulence \citep{Porth2015,Singh2016,Bromberg2016}. Magnetic reconnection and turbulence can then accelerate nonthermal particles \citep{Alves2018,Medina2020,Nishikawa2020,Davelaar2020}. Since the striped jet model has a strong toroidal magnetic field component, thus we expect that both kink instabilities and large-scale reconnection between stripes can co-exist in this picture. Generally speaking, kink instabilities mostly affect the central spine of the jet without disrupting the jet geometry \citep{ONeill2012,McKinney2009}. Therefore, most of the energy dissipation and particle acceleration, and the resulting synchrotron emission, concentrate in the central spine. However, observations suggest that the jet can be opaque to the radio emission due to the strong synchrotron self absorption. Therefore, the radio emission from the central spine may not account for the observed radio signals. However, if there is a significant flattening in the external density profile, \citet{Barniol2017} have demonstrated that strong kink instabilities can be triggered, which can affect a large cross section of the jet and lead to significant magnetic energy dissipation. This flattening may originate from that the jet propagates out of the accretion flow region, which is likely at several parsec-scales from the central engine. Recently, \citet{Dong2020} have found that the strong energy dissipation and radiation concentrate near the density profile flattening at $\sim $ parsec from the central engine, which can explain the blazar zone radiation signatures. However, the radial extent of the dissipation region is, in this scenario, limited, making it unclear how the kink-induced magnetic reconnection can explain the large-scale jet radiation. Additionally, in a considerably magnetized environment ($\sigma > 1)$, \citet{Davelaar2020} have shown that the current sheets produced by kink instabilities often have a strong guide field component, which is the magnetic component that is perpendicular to the reconnecting components. The strong guide field may result in relatively slow reconnection and rather steep particle spectrum \citep{Werner2017,Guo2020}. On the other hand, in a moderately magnetized environment where $\sigma \sim 1$, similar to the peak dissipation region of the striped jet model, recent works have shown efficient particle acceleration driven by both reconnection and turbulence in the kinked region \citep{Medina2020,Nishikawa2020,Kadowaki2020}. This can lead to overall similar radiation signatures as in the striped jet model. Additional constraints through, for instance, polarization signatures, may help to distinguish these models \citep{Zhang2019r,Zhang2020}.

Our analytical study of the striped jet observable signatures is based on a steady conical jet model. In a more realistic situation, kink instabilities and recollimation shocks may change the geometrical appearance of the jet. Since the striped jet model predicts a broad energy dissipation profile in distance, if such changes in the jet geometry happen before the peak of the dissipation, they may modify the dissipation profile. For example, recollimation shocks can happen at various scales from the central engine from sub-parsec to several parsecs, depending on the external density profile. Those shocks on sub-parsec scales may squeeze and slow down the jet, affecting its reconnection timescale. On the other hand, strong kink instabilities as shown in \citet{Barniol2017} are likely to happen at several to tens of parsecs from the central engine. They are beyond the peak of the dissipation profile under the striped jet model, where the magnetization factor also drops below unity. Therefore, their effects on the striped jet dissipation profile are expected to be minor. Moreover, the blazar jet is highly dynamical. Observations have shown moving radio knots and outburst of new radio components from the unresolved radio core \citep{Marscher2008,Jorstad2017}. These time-dependent features are beyond our analytical model. We expect that these additional features can considerably complicate the identification of spectral luminosity peak $\nu_{\rm peak}$ and its core-shift distance of the large-scale jet emission, which are crucial to locate the blazar zone and probe the black hole accretion.

\section*{Data Availability}
The data underlying this article will be shared on reasonable request to the corresponding author.

\section*{Acknowledgements}

We thank the anonymous referee for the very helpful review. HZ and DG acknowledge support from the NASA ATP NNX17AG21G, the NSF AST-1910451, the NSF AST-1816136 grants and  by Fermi Cycle 12 Guest Investigator Program \#121077.




\bibliographystyle{mnras}
\bibliography{Stripe_Jet_Blazar} 








\bsp	
\label{lastpage}
\end{document}